# Improving Students' Understanding of Quantum Measurement

Guangtian Zhu and Chandralekha Singh

*Department of Physics and Astronomy, University of Pittsburgh, Pittsburgh, PA, 15260, USA*

**Abstract.** We describe the difficulties advanced undergraduate and graduate students have with quantum measurement. To reduce these difficulties, we have developed research-based learning tools such as the Quantum Interactive Learning Tutorial (QuILT) and peer instruction tools. A preliminary evaluation shows that these learning tools are effective in improving students' understanding of concepts related to quantum measurement.

**Keywords:** Quantum mechanics, measurement, tutorials, peer instruction tools

## INTRODUCTION

There have been many investigations of the difficulties students have in learning quantum mechanics (QM) [1-4]. Based upon the findings, we are developing a set of research-based learning tools including Quantum Interactive Learning Tutorials (QuILTs) and concept tests similar to those developed earlier for introductory physics courses [4,5]. The QuILTs use a guided inquiry-based approach to learning QM. They also assist students in building a knowledge structure by helping them to discern the coherence in the framework of QM. The concept tests are integrated with lectures and encourage students to learn from each other.

In this paper, we discuss the investigation of students' difficulties with quantum measurement. We build upon this research to design and evaluate research-based learning tools to help students develop a good grasp of the quantum measurement formalism. The investigation of students' difficulties was conducted with several hundred undergraduate and graduate students at the University of Pittsburgh (Pitt) and other universities by administering written tests and conducting in-depth individual interviews with a subset of students.

The research-based QuILT and concept tests related to quantum measurement were administered to students in the first semester of a full-year junior-senior level QM course. To assess the effectiveness of the QuILT and concept tests on quantum measurement, we gave the same assessment related to quantum measurement to the experimental and comparison groups in different but equivalent classes at two similar universities. The comparison group only had traditional lectures and weekly homework in a similar two-semester QM class in which the same textbook was used. Our prior investigation shows that the students' performance on tests given in the upper-level QM courses at the two universities was comparable when traditional instruction was used at both institutions [3].

## INVESTIGATION OF DIFFICULTIES

Our goal was to examine students' knowledge of quantum measurement after traditional instruction. To simplify the mathematics and focus on the concepts related to quantum measurement, we often used a one-dimensional (1D) infinite square well model. Both open-ended questions and multiple choice questions were administered to probe students' difficulties.

We find that many students were unclear about the difference between energy eigenstates and eigenstates of operators corresponding to the other physical observables. They were also unclear about what happens to the state of the system after the measurement of an observable. Many students struggled to distinguish between the measured value, the probability of measuring it and the expectation value. Students were also confused about whether the system is "stuck" in the state in which it collapsed right after the measurement or whether it reverts back to the state prior to the measurement. Here, we will only elaborate on students' difficulties with the time evolution of the wavefunction after a quantum measurement.

Within the Copenhagen interpretation of QM, the measurement of an observable is treated separately from the "normal" time-evolution of the system according to the Time Dependent Schroedinger Equation (TDSE). When a measurement of an observable is performed, the state of the system instantaneously collapses to an eigenstate of the corresponding operator after which the system will evolve according to the TDSE. To investigate the difficulties with the time development of the wavefunction, one question was about consecutive position measurements for a 1D infinite square well:

*Q0. If you make a measurement of position on an electron in the ground state and wait for a long time before making a second measurement of position, do you expect the outcome to be the same in the two measurements? Explain.*

To answer this question correctly, students must know the following: (1) The ground state wavefunction will collapse to a position eigenfunction (a delta function) after the first position measurement. (2) The position eigenfunction is a non-stationary state wavefunction so it will evolve in time in a non-trivial manner and the system will not in general be found in a position eigenstate at a time $t$. Thus, after a long time, the second measurement of position in general will yield a different value from the first measurement. Students had the following common difficulties:

**Difficulty 1: The system remains in the ground state after a position measurement**

In response to the consecutive position measurement question Q0, some students stated that the system will be in the ground state after both the first and second position measurements. Interviews suggest that the students making these types of responses often did not realize the difference between an energy eigenstate and a position eigenstate.

In a multiple-choice survey administered to 76 students from six universities, students were asked about a situation where the position of the particle in the state $(\psi_1 + \psi_2)/\sqrt{2}$ (superposition of the ground and first excited states) is measured first and then the energy is measured immediately after that. Forty-nine percent of the students incorrectly claimed they could only measure energies $E_1$ or $E_2$. Individual discussions suggest that students believed the system remains in the initial state after the position measurement.

**Difficulty 2: The system remains in a position eigenstate at all times after a position measurement**

On the other hand, some students thought that after the first position measurement, the system gets "stuck" in a position eigenstate. They did not know that the position eigenfunction evolves in time in a non-trivial manner and the system does not remain in a position eigenfunction for all future times $t$. These students stated that the second position measurement will give the same value as the first one unless there was an "outside disturbance". In the multiple-choice survey, 23% of the 76 students incorrectly believed that if a particle has a definite value of position at time $t=0$, the position of the particle is always well-defined for $t>0$.

**Difficulty 3: The system returns to the initial state**

Students were also asked a series of questions related to measurement when the initial state of the system is $\sqrt{2/7}\psi_1(x) + \sqrt{5/7}\psi_2(x)$ for an electron confined in a 1D infinite square well of width $a$ as follows:

*Q1. If the energy measurement yields $4\pi^2\hbar^2/2ma^2$, what is the wavefunction right after the measurement?*

*Q2. Immediately after the energy measurement in Q1, you measure the position of the electron. What possible values could you obtain and what is the probability of each?*

*Q3. After the position measurement in Q2, you wait for time t and measure the position again. Would the probability of measuring each possible value be different from Q2?*

When Q1 was given in the multiple-choice format to 76 students who were asked about the state of the system long after the energy measurement; 20% of the students incorrectly thought that the wavefunction would remain in the original state after the measurement; 36% of the students claimed that the wavefunction will collapse upon the energy measurement but evolve back to the initial state long time after the measurement. During the individual interview, a student said. "…it's like tossing a coin. You can get either head or tail after the measurement. But when you make another measurement, it goes back to a coin (with two sides)."

The following question about the measurement of position also revealed difficulties about the measured state going back to the state before measurement:

*You perform a position measurement of the particle in a finite square well in the first excited state. Choose all of the following statements that are correct:*
*(1) Right after the position measurement, the wavefunction will be peaked about a particular value of position.*
*(2) A long time after the position measurement, the wavefunction will go back to the first excited state wavefunction.*
*(3) The wavefunction will not go back to the first excited state wavefunction even if you wait for a long time after the position measurement.*

Seventy-three percent of the 76 students knew that the wavefunction would collapse to a position eigenstate after the measurement. However, only 26% of the students correctly answered this question by choosing both (1) and (3). Forty-one percent believed that the wavefunction will return to the state before measurement after a long time.

**Difficulty 4: Use of classical concepts to analyze the time evolution of a quantum system**

When answering Q2, some students had difficulty in differentiating between the probability of measuring position and the expectation value of position. Moreover, none of the students after the traditional instruction provided a completely correct response to Q3 which assessed the same concepts as in question Q0 discussed earlier. Some students used a classical description for the time-evolution after the measurement of position as in the following response: "the electron moves around". Individual discussions suggest that such responses reflect the difficulty in visualizing and describing the time evolution of the wavefunction of a quantum system via the TDSE.

## IMPROVING STUDENT LEARNING

The goal of the measurement QuILT is to build connections between the formalism and conceptual aspects of quantum measurement without compromising the technical aspects [4]. The QuILT builds on the prior knowledge of students and was developed taking into account the difficulties found in the written surveys and think-aloud interviews. It uses computer-based visualization tools from Open Source Physics [6] to help students build a physical intuition about concepts related to the quantum measurement. The QuILT development went through a cyclical iterative process which includes the following stages: (1) Development of the preliminary version based upon theoretical analysis of the underlying knowledge structure and research on students' difficulties, (2) Implementation and evaluation by administering it individually to students, measuring its impact on student learning and assessing what difficulties remained, (3) refinement and modification based upon the feedback from the implementation and evaluation.

One effective strategy to help students build a robust knowledge structure is to induce "cognitive conflict" in students' minds so the students realize that there is some inconsistency in their reasoning, and then provide them appropriate guidance and support. In the measurement QuILT, after predicting what they expect in various situations, students are asked to check their predictions using simulations. If the prediction and observations do not match, students reach a state of cognitive conflict. At that point the QuILT provides them guidance to reconcile the difference between their predictions and observations so that they can build a good grasp of relevant concepts.

Before working on the QuILT, students are provided with a set of warm-up activities to help them review the necessary background knowledge about quantum measurement. The QuILT itself consists of two parts: (1) measurement outcomes and their probabilities, and (2) the time evolution of the system after a measurement. Here, we briefly discuss only the section in the QuILT related to the time evolution after the measurement of the position of the particle.

The measurement QuILT helps students with issues related to the position measurement. Students predict theoretically what state they could obtain after a position measurement and then they use the simulation to check their prediction. In an ideal position measurement, the state of the system would collapse to a delta function at a position where the probability of measuring the position is non-zero. As shown in Fig. 1, the initial ground state collapses to a peaked Gaussian packet due to the computational limitations in constructing a very peaked function. However, the QuILT uses this opportunity to help students recognize that a delta function is a theoretical construct and the position measurement in the real world situations, e.g., in a double slit experiment, where single particles land on the screen after passing through the slit, will have an uncertainty in position.

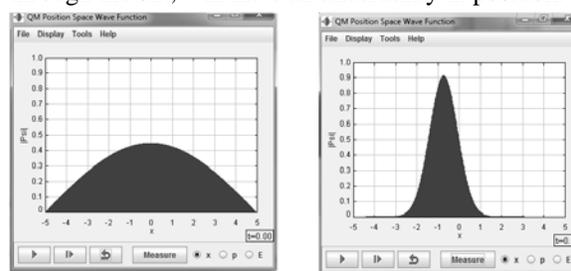

(a) before measurement    (b) after measurement
**Fig 1.** Position measurement in an energy eigenstate.

After predicting what should happen if position measurements are performed on a large number of identically prepared systems, students are asked to reset the initial state of the system in the simulation and repeat the position measurement. They observe that the center position of the collapsed wavefunction is generally different but its shape is always the same. They verify this result in multiple contexts, e.g., for different quantum systems and different initial states.

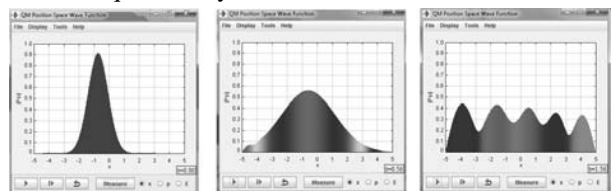

*(a)* time *t=0 units*   *(b) t=0.56 units*   *(c) t=1.56 units*
**Fig 2.** Time evolution of the position eigenfunction.

As noted earlier, many students had the misconception that, after the position measurement, the position eigenfunction does not change with time. In the QuILT, students are asked to use the simulation after their initial prediction for what should happen as a function of time after they perform a position measurement. In an ideal measurement, at the instant the position is measured, the wavefunction of the system will collapse to a delta function $\delta(x-x_0)$ and the measured position will be $x_0$. But in the simulation, it collapses to a peaked wavefunction as shown in Fig. 2(a) which is a linear superposition of a finite number of energy eigenstates. Different energy eigenstates will have their own time-dependent phase factors and the wavefunction would not be peaked except at some special times for the 1D infinite square well (due to the periodic behavior of the time evolution of the wavefunction). Figs. 2(b) and (c) show two snapshots.

Apart from learning the pictorial representation in the simulations, the QuILT helps students interpret the time evolution of the position eigenfunction via the TDSE and discern the central role of the Hamiltonian of the system in the evolution. The following is an example of such a guiding question:

*Given the wavefunction at time t=0, why is it useful to write the state of a quantum system as a superposition of energy eigenstates to find the wavefunction after time t?*

Students must realize that the Hamiltonian governs the time evolution of the system according to the TDSE so the eigenstates of the Hamiltonian are special for issues related to the time evolution of the wavefunction. Help is provided at the end of the QuILT if students are struggling with these issues.

We find that despite learning that a position eigenstate is not a stationary state, some students still have the misconception that after a position measurement, the position eigenstate would finally return to the initial state *before* the measurement. After the prediction phase, the simulations are helpful in reducing this difficulty. As shown in Fig. 2, the students observe that the delta function does not return to the initial state (e.g., the ground state in Fig. 1). Students also perform a systematic mathematical analysis of the time-dependence of the wavefunction in the superposition of energy eigenstates to convince themselves that a position eigenstate cannot go back to the state before the measurement of position.

In addition to the QuILT, research-based sequence of concept tests developed using an iterative process can be integrated with lectures to improve students' understanding. Students reflect on them with peers [5].

## PRELIMINARY EVALUATION

We designed two equivalent versions of a test to assess student learning. For the experimental group, if Test A was given to a student after the administration of measurement related concept tests, then Test B was given after the student had also worked on the measurement QuILT and vice versa. In the comparison group with traditional instruction, 15 students were randomly given Test A and 10 students were given Test B. In the experimental group, 6 students were randomly given Test A and 7 students took Test B after instruction that included concept tests. The experimental group was given the version of the test they had not attempted earlier after they learned about measurement using *both* the concept tests and QuILT.

The class average in the comparison group with traditional lectures was 26% including both versions (Tests A and B). Students in the experimental group had been using the concept tests as a peer instruction tool in class since the first day of the semester. The first test was given to these students after relevant lectures with the concept tests and the average score was 68%. Then, the experimental group students worked on the QuILT in class and their average score including both versions after the QuILT was 91%.

Table 1 only lists the students' performance for the questions Q0 in Test A and Q1, Q2, Q3 in Test B, as discussed earlier. The analysis of students' difficulties with these questions is summarized in the section "Investigation of Difficulties". Since different number of students had taken Tests A and B, the number of students answering each question is shown in the parentheses in Table 1.

|    | Comparison | Concept test | QuILT |
|----|------------|--------------|-------|
| Q0 | 37% (15)   | 71% (6)      | 100% (7) |
| Q1 | 35% (10)   | 86% (7)      | 100% (5) |
| Q2 | 10% (10)   | 36% (7)      | 90% (5) |
| Q3 | 5% (10)    | 64% (7)      | 100% (5) |

**Table 1.** Percentage of correct responses to questions about quantum measurement by different groups of students

## SUMMARY

Students struggle with issues related to the time evolution of the wavefunction after measurement. We developed a research-based QuILT and concept tests to improve students' understanding of quantum measurement concepts. Both these learning tools keep students actively engaged in the learning process. They provide a guided approach to bridge the gap between the quantitative and conceptual issues related to quantum measurement, help students connect different concepts and build a knowledge structure. Our preliminary results show that these learning tools significantly improve students' understanding.

## ACKNOWLEDGMENTS

We thank the National Science Foundation.

## REFERENCES

1. P. Jolly, D. Zollman, S. Rebello, and A. Dimitrova, Am. J. Phys. 66 (1), 57-63 (1998).
2. M. Wittmann, R. Steinberg, E. Redish, Am. J. Phys. 70 (3), 218-226 (2002).
3. C. Singh, Am. J. Phys. 76 (3), 277-287 (2008).
4. C. Singh, Am. J. Phys. 76 (4), 400-405 (2008).
5. E. Mazur, Peer Instruction, Prentice Hall, (1997).
6. For example, see http://www.opensourcephsyics.org